\documentclass[10pt,a4paper]{article}
\usepackage{graphicx} 

\usepackage{hyperref}
\usepackage{amsmath}
\usepackage{amsfonts}
\usepackage[sort&compress,numbers]{natbib}
\usepackage{notes2bib}
\usepackage{graphicx, caption, subcaption}
\usepackage[left=2.5cm, right=2.5cm]{geometry}
\usepackage{times}
\usepackage{titlesec}
\usepackage{ragged2e}
\usepackage{comment}
\usepackage{multirow}
\usepackage[titletoc]{appendix}
\begin{document}


\thispagestyle{plain}

\thispagestyle{plain}
\begin{center}
    \Large
    \textbf{Understanding the rift between update rules in Evolutionary Graph Theory: The intrinsic death rate drives star graphs from amplifying to suppressing natural selection}
        
    \vspace{0.4cm}
    \normalsize
    \textbf{Max Dew$^{1,\ast}$ Christopher E. Overton$^{1,2,\dagger}$} \\

    $^1$\textit{Department of Mathematical Sciences, University of Liverpool, Liverpool, UK} \\
    $^2$\textit{Modelling Division, Analysis and Intelligence Assessment, Chief Data Officer Group, UK Health Security Agency} \\
    \vspace{0.9cm}
    \textbf{Abstract} 
\end{center}
Evolutionary graph theory is the study of evolutionary dynamics in structured populations. A well-known problem in evolutionary graph theory is that the spread of mutation (measured by fixation probability) is impacted by the graph type chosen and the update rule. For example, the star graph is an amplifier of natural selection under the birth-death with fitness on birth (Bd) update rule but a suppressor of natural selection under the death-birth with fitness on birth (dB) update rule. A continuous-time EGT model has been found to replicate Bd and dB results as special cases. Using this model, we show that changing the natural (intrinsic) death rate can cause a shift from Bd to dB dynamics. Assuming the mutant is advantageous, we show that if the natural death rate is greater than $\frac{1}{\sqrt{N}}$ the star is a suppressor, where $N$ is the number of nodes. As $N \longrightarrow \infty$, the natural death rate required to drive the star to a suppressor tends towards zero, so as the size of the graph increases, the star graph is likely to be suppressing for any non-zero natural death rate. 
\vspace{9.0cm} \\
$\ast$ Corresponding author: \href{mailto:M.Dew@liverpool.ac.uk}{M.Dew@liverpool.ac.uk}, Orchid ID: \url{https://orcid.org/0009-0009-7661-7369} \\
$\dagger$ Orchid ID: \url{https://orcid.org/0000-0002-8433-4010} \\

\clearpage \newpage 

\section{Introduction} \label{sec:introduction}
Evolution describes the process by which species adapt and change over time. The study of evolution is not only of interest due to scientific curiosity, but also allows the development of new techniques to make evolutionary predictions or influence evolution (evolutionary control) \cite{wortel2023towards}, for example, predicting the spread of infectious diseases \cite{InfectiousDiseasesStructureRealWorld}.  A key question in the study of evolution is that if a single mutant arises in a population of identical residents, what is the probability that the mutant takes over the resident population (known as  \textbf{fixation probability} \cite{patwa2008fixation,adlam2015Amplifiersofselection,EvolutionaryDynamicsBook,ReviewGraphGameTheoryApplication})? One model to describe this kind of evolutionary process is the \textbf{Moran process}, first introduced in 1958 by Moran (1958) \cite{moran1958random} and further developed in Moran (1962) \cite{moran1962statistical}.  The Moran process considers a finite haploid population consisting of two types of individuals, mutants and residents, who experience birth or death events \cite{moran1958random}. The population is assumed to be homogeneous or well-mixed - every individual interacts with every other individual equally \cite{DynamicsStructuredPopulations}. However, real populations are not necessarily well-mixed but rather heterogeneous \cite{DynamicsStructuredPopulations}. For example, some microbiology experiments specifically study heterogeneous populations where the microbe populations are not shaken or well aerated, resulting in substructures \cite{mcdonald2019microbial}.  While it is assumed that shaken aerated cultures are well-mixed \cite{mcdonald2019microbial}, some research has shown that this might not be the case \cite{marrec2021toward}. 
\newline \newline 
Nowak et al. (2005) \cite{EvolutionaryDynamicsGraphs} extended the Moran process by considering heterogeneous populations (populations with a structure) represented by a graph \cite{ReviewGraphGameTheoryApplication}, creating a new field known as Evolutionary Graph Theory (EGT). The individuals exist on the nodes (or vertices) \cite{EvolutionaryDynamicsBook,ReviewGraphGameTheoryApplication}, one individual per node \cite{ManyUpdatesDirectedUndirected}, and the edges represent an interaction between individuals \cite{herrerias2019motion,DynamicsStructuredPopulations,RandomGraphsBDDynamics}. With the introduction of heterogeneity (or structure) in the population, how does this impact the spread of mutation within this population? By comparing how the fixation probability changes relative to that of a homogeneous population, it is possible to see if the structure amplifies or suppresses the spread of mutation \cite{adlam2015Amplifiersofselection,DynamicsStructuredPopulations}. Nowak et al. (2005) \cite{EvolutionaryDynamicsGraphs} provided some examples, showing that the star graph is an \textbf{amplifier} of selection (increases spread of mutation), the burst is a \textbf{suppressor} of selection (decreases the spread of mutation) and the complete graph provides the same result as a homogeneous population from the Moran model. 
\newline \newline 
A problem that has arisen when determining the impact of population structure is that changing the update rule, i.e. the mechanism by which birth and death take place, can lead to qualitative differences on whether a certain graph is an amplifier or suppressor~\cite{yagoobi2023categorizing}. For example, for an update of birth-death with fitness on birth (Bd)~\cite{EvolutionaryDynamicsGraphs} a star is an amplifier, but for death-birth with fitness on birth (dB)~\cite{SimpleRuleEvolutionCooperation}, a star is a suppressor \cite{RandomGraphsBDDynamics}. 
The aim of this project is to understand the biological drivers behind this qualitative change. We will discuss the known problem of the update rules with the classical EGT model, which we will refer to as discrete-time EGT model. We will then discuss a biologically motivated eco-evolutionary model  \cite{EGTMarkovContinuous} that is
able to recreate the results from the discrete-time EGT model by suppressing ecological dynamics \cite{EGTMarkovContinuous}, which we will refer to as continuous-time EGT model. Using the continuous-time EGT model, we will show that it is the natural death rate of individuals that drives the change from an amplifier to a suppressor. Focusing on the star graph as a well-known amplifier in Bd, we find an upper bound for this change in star graphs and show that, for very large systems, if the natural death rate is not zero, a star would be a suppressor.  

\section{Methods}
\subsection{Graphs} \label{sec:graphs_definition}
Within evolutionary graph theory, we consider graphs defined by a right stochastic matrix $W$, where the element of the matrix $w_{ij}$ \cite{EvolutionaryDynamicsBook} is the weight of the edge from node $j$ to node $i$~\cite{StarFixationProbabilityExact}, representing the strength of interaction between the two nodes \cite{herrerias2019motion,DynamicsStructuredPopulations,RandomGraphsBDDynamics}, defined as \cite{EvolutionaryDynamicsBook}
\begin{equation}
    W = 
    \begin{cases} 
        w_{ij} & \text{if there is an edge between } j \text{ and } i, \\
        0 & \text{otherwise.} \\
    \end{cases}
\end{equation}
The nodes usually represent individuals \cite{herrerias2019motion}, but can also represent patches~\cite{pattni2023eco}, similar to meta-population models \cite{yagoobi2023categorizing,EGTMarkovContinuous,yagoobi2021fixation}.
\newline \newline 
An undirected graph is where the edges are bidirectional \cite{RandomGraphsBDDynamics} ($w_{ij} >0 \text{ and } w_{ji} > 0$ \cite{allen2020transient}) and unweighted if all edges leaving a node have equal weight \cite{StarFixationProbabilityExact}. The graphs contain self-loops  if the node is connected to itself \cite{RandomGraphsBDDynamics} ($w_{ii} \neq 0$ \cite{allen2020transient}). Only undirected unweighted graphs with self-loops are considered in this paper. 

\subsubsection{Complete graph} \label{sec:graphs_definition_complete}
A \textbf{complete} graph is a representation of homogeneous \cite{DynamicsStructuredPopulations} or well-mixed populations \cite{adlam2015Amplifiersofselection,RandomGraphsBDDynamics}, and is defined as a fully connected graph \cite{StarUpdatesQualitativelyExplained}, where all the nodes are connected to every other node with equal weights. \cite{DynamicsStructuredPopulations} The elements $w_{ij}$
for a complete graph of $N$ nodes with self-loops are defined as \cite{EGTMarkovContinuous}
\begin{equation}
            w_{ij} = 1/N \text{ for all $i$ and $j$.} 
\end{equation}

\subsubsection{Star graph} \label{sec:graphs_definition_star}
A star graph with self-loops consists of two kinds of nodes - a central node, which connects to every node (including itself), and a leaf node, which connects only to the central node and itself~\cite{EGTMarkovContinuous,adlam2015Amplifiersofselection}. The elements $w_{ij}$ 
for a star graph of $N$ nodes with self-loops are defined as \cite{EGTMarkovContinuous}
\begin{equation}
    w_{ij} = 
    \begin{cases} 
        1/N & \text{if $i=n$ for all $j$} \\
        1/2 & \text{if $j=n $ for all $i\neq n$ }, \\
        1/2 & \text{if $i\neq n $ } \\
        0 & \text{otherwise,} \\
    \end{cases}
\end{equation}
where $n$ is the index of the central node.

\subsection{Definitions}
\subsubsection{Fixation probability} \label{sec:def_fixation_probability}
Within evolutionary modelling, we are interested in the dynamics of a rare mutant introduced into a population of residents when the mutation rates of individuals are equal to zero \cite{moran1958random}. A key quantity to measure these dynamics is the \textbf{fixation probability}. The fixation probability of a mutant is the probability of a mutant individual taking over a population of residents \cite{EvolutionaryDynamicsBook,ReviewGraphGameTheoryApplication}. The opposite of fixation is extinction, where only the residents remain~\cite{ReviewGraphGameTheoryApplication}. In this paper, we only consider fixation probability of a single mutant in a population of $N-1$ residents.

\subsubsection{Amplifiers and suppressors} \label{sec:def_amplifiers_suppressors}
Let $\rho(\beta_1)$ denote the fixation probability of a mutant of fitness $\beta_1$. Resident fitness is defined as $\beta_0$. To determine if a graph is an amplifier or suppressor, the fixation probability of that graph is compared to the fixation probability of a homogeneous population $\rho_{\text{homogeneous}}$, which acts as the baseline population structure \cite{yagoobi2023categorizing,bhaumik2024constant,alcalde2018evolutionary}. A mutant is considered advantageous when $\beta_1>\beta_0$ \cite{adlam2015Amplifiersofselection,allen2020transient}. When $\beta_1<\beta_0$ the mutants are called deleterious \cite{allen2020transient}, and when $\beta_1=\beta_0$ this is referred to as neutral drift as evolution does not play a part \cite{allen2020transient,ReviewGraphGameTheoryApplication}. 
\newline \newline 
Amplifiers are defined as \cite{allen2020transient,adlam2015Amplifiersofselection}
\begin{equation} \label{eq:amplifiers_definition_maintext}
\text{Amplifiers = }
    \begin{cases}
        \rho(\beta_1) < \rho_{\text{homogeneous}} \text{ for } 0 < \beta_1 < \beta_0 \\ 
        \rho(\beta_1) > \rho_{\text{homogeneous}} \text{ for } \beta_1>\beta_0 \text{.}
    \end{cases}
\end{equation}

\noindent Suppressors are defined as \cite{allen2020transient,adlam2015Amplifiersofselection,alcalde2018evolutionary} 
\begin{equation} \label{eq:suppressors_definition_maintext}
\text{Suppressors = }
    \begin{cases}
        \rho(\beta_1) > \rho_{\text{homogeneous}} \text{ for } 0 < \beta_1 < \beta_0 \\ 
        \rho(\beta_1) < \rho_{\text{homogeneous}} \text{ for } \beta_1>\beta_0 \text{.}
    \end{cases}
\end{equation}
This works considers only advantageous mutants, so $\beta_1>\beta_0$.  

\subsection{Discrete-time EGT model} \label{sec:discrete_time_model}
To demonstrate the issue with the choice of update rule leading to qualitative differences on whether a certain graph is an amplifier or suppressor, we first introduce the classical discrete-time EGT model, as described widely in the literature \cite{RandomGraphsBDDynamics,EvolutionaryDynamicsGraphs,ManyUpdatesDirectedUndirected,StarUpdatesQualitativelyExplained,DynamicsStructuredPopulations,ReviewGraphGameTheoryApplication}. The population of individuals exist on the nodes of the graph, one individual per node. At the initial state of the system, the population consists of one mutant that is placed uniformly at random \cite{RandomGraphsBDDynamics} (uniform initialisation \cite{adlam2015Amplifiersofselection, marrec2021toward}) and the rest are residents \cite{RandomGraphsBDDynamics}. At each time-step, an update occurs consisting of a birth event where an offspring is produced (assumed identical to the parent) and a death event where an individual dies \cite{RandomGraphsBDDynamics}, the order of which depends on the choice of update rule~\cite{yagoobi2023categorizing, ReviewGraphGameTheoryApplication}. The first event is a global event where an individual from the entire graph is chosen. The second event is a local event where a neighbour of the chosen individual is selected~\cite{RandomGraphsBDDynamics}. 
The result of the update is that the offspring is placed on the empty node where the individual died \cite{starfixationfurtherdetail,pattni2015evolutionary,RandomGraphsBDDynamics}. The end states are a population of either only mutants or only residents,
with the other type of individual becoming extinct \cite{ReviewGraphGameTheoryApplication}. 
\newline \newline 
Multiple update rules have been identified \cite{pattni2015evolutionary, yagoobi2023categorizing,sharma2025graph,ReviewGraphGameTheoryApplication}, and the two update rules that we focus on are Birth-Death with fitness on birth (Bd) and Death-Birth with fitness on birth (dB). Both update rules relate the fitness of an individual to their birth rate, assuming fitness is proportional to the birth rate or the reproductive success of an individual. The mutant has a fitness of $\beta_1$ and the resident has fitness of $\beta_0$, with the baseline typically set to $\beta_0=1$ \cite{RandomGraphsBDDynamics,EvolutionaryDynamicsGraphs,alcalde2018evolutionary}. For Bd, the birth event occurs first with an individual from the entire graph selected to reproduce (global event) with probability proportional to its fitness. This is then followed by the death event, with an individual selected from the neighbours of the birthing node to die (local event) uniformly at random. The mutant and resident are assumed equally likely to die \cite{StarUpdatesQualitativelyExplained,ManyUpdatesDirectedUndirected,DynamicsStructuredPopulations}. In dB, the opposite occurs with the death event being a global event (uniformly at random) and the birth event being local (with probability proportional to fitness)~\cite{RandomGraphsBDDynamics}.
\newline \newline 
It has been found that the choice of the update-rule impacts whether a structure is an amplifier or suppressor~\cite{yagoobi2023categorizing}. For example, a star has been identified as an amplifier for Bd but a suppressor for dB \cite{RandomGraphsBDDynamics,ManyUpdatesDirectedUndirected,StarUpdatesQualitativelyExplained}. Further discussions of this can be found in Section~\ref{results:the_problem_discrete_model}. In most cases, the choice of update rule is arbitrary, and it is not clear which biological properties influence this~\cite{herrerias2019motion}. Therefore, it is hard to make generalisations about the impact of population structure on the evolution of real populations. 

\subsection{Continuous-time EGT model} \label{sec:continuous_time_model}
It is clear there are some assumptions in the discrete-time EGT model. The population size is assumed to be constant (ecological equilibrium \cite{EGTMarkovContinuous}), which is achieved by the birth and death events occurring at the same time-step, and the two events are coupled \cite{herrerias2019motion}. The motivation behind an eco-evolutionary model developed by Pattni et al. (2021) \cite{EGTMarkovContinuous} was to decouple birth and death events. This was based on Champagnat (2006) \cite{champagnat2006unifying} with the addition of graph structure. As special cases, this model has been shown to yield Bd and dB dynamics \cite{EGTMarkovContinuous}. The possible events are birth with mutation, birth without mutation, and death. In this paper we only consider the latter two. The ecological aspect of the model allows meta-populations - more than one individual can occupy a node at a time - and allows variable population size. At each time-step, only one event occurs, either a birth event or a death event \cite{EGTMarkovContinuous}. As we are interested in understanding the rift between Bd and dB in the discrete-time EGT model, we consider the special case of a fixed population size with one individual per node. 
\begin{table}[h!]
\centering
\begin{tabular}{ |p{2cm}|p{10.5cm}|p{2cm}|  }
 \hline
 \multicolumn{3}{|c|}{Parameters list} \\
 \hline
 Parameters& Description&Death/Birth\\
 \hline 
 $\gamma$&competition rates, death rate due to competition between individuals&Death\\ 
 \hline
 $\delta$&natural death rate of the individual/intrinsic mortality rate, death rate not caused by external factors&Death\\
 \hline
 $\beta$&birth rate of the individual&Birth \\
 \hline
 $w_{n_i,n_j}$&representing the weight of the edge between node $n_i$ and $n_j$, the strength of the connection between node $n_i$ and $n_j$&Birth\\
 \hline
 $x_{n_j}$&the number of individuals existing on the node $n_j$&Birth\\
 \hline 
 $s$&the offspring survival rate, the likelihood of the offspring survival that is impacted by $x_{n_j}$&Birth\\
 \hline 
 $c$&amplifier factor&Both\\
 \hline
\end{tabular}
\caption{Parameters definitions of the continuous-time EGT model and whether they relate to birth or death event}
\label{table:parameters_EGTcontinous}
\end{table}
\newline \newline 
The parameters are defined in Table \ref{table:parameters_EGTcontinous} alongside whether they relate to the birth or death event. The offspring survival rate $s$ is linked to $x_{n_j}$ and is defined between $0$ and $1$. If $s=0$, then offspring can only survive on unoccupied nodes ($x_{n_j}=0$). If $s=1$, then the offspring will survive on the node no matter the number of current occupants \cite{EGTMarkovContinuous}. 
\newline \newline 
The continuous-time EGT model can be used to recreate the results of the discrete-time EGT model by suppressing the ecological dynamics with a feedback loop \cite{EGTMarkovContinuous}. This is achieved by amplifying the birth and death rates such that the difference between the time-steps is small and the birth and death events interchange at each time step (i.e. a birth event is always followed by a death event and a death event is always followed by a birth event). This is implemented through an amplifier factor $c$ in the following way \cite{EGTMarkovContinuous}:   
\begin{itemize}
    \item[1)]If there is more than one individual on the node, their death rate is amplified resulting in the death event occurring next and so only one individual remains on the node.
    \item[2)] If there is an empty node, the birth rate of neighbouring individuals is amplified, so the birth event occurs next, and the offspring is placed on the empty node.
\end{itemize}
 The convergence to the discrete-time EGT model is shown to be guaranteed at the limit of $c \longrightarrow \infty$~\cite{EGTMarkovContinuous}. For numerical studies, $c$ is chosen to be sufficiently large that the results between the two models are negligible \cite{EGTMarkovContinuous}. To be consistent with the discrete-time EGT model, we also have the same initial set-up, one individual occupies each node, with one initial mutant that is randomly placed on the graph and the rest are residents \cite{EGTMarkovContinuous}. 
\newline \newline 
The birth rate is defined in terms of an individual's fitness. The natural death rate $\delta$ is the same for both mutants and residents and the competition rates $\gamma$ are the same for all possible interactions between mutants and residents. The values of natural death rate $\delta$ and the offspring survival rate $s$ determine if the update is  Bd or dB. This can be seen from the replacement rate for an individual of type $u$ on node $n_i$  by an offspring produced by an individual of type $v$ who is on $n_j$, defined as \cite{EGTMarkovContinuous}
\begin{equation} \label{eq:replacement_rate_sim_maintext} 
    r(u,v,n_i,n_j) = s \left( \beta_{u} w_{n_i,n_j} \frac{1}{2} \right) + \delta \left( \frac{\beta_{u} w_{n_i,n_j}}{\sum_{0}^{k \setminus \{j\}} \beta_{k} w_{n_k,n_j}} \right) \text{.}
\end{equation}
Note that, in this setup, $\gamma$ no longer plays a role in the replacement rate and hence the overall dynamics of the system. 
\newline \newline 
Looking at equation \ref{eq:replacement_rate_sim_maintext}, it becomes clear why the parameters $\delta$ and $s$ are those that determine if the dynamics of the system are Bd or dB. As proven in Pattni et al. (2021) \cite{EGTMarkovContinuous}, if $\delta=0$ and $s=1$ then we get Bd dynamics and if $\delta=1$ and $s=0$ then we get dB dynamics. 
This means that Bd and dB dynamics can be recreated as special cases of the continuous-time EGT model \cite{EGTMarkovContinuous}. 
\newline \newline
Within this model, expressions for the fixation probability on complete and star graphs have been defined as \cite{EGTMarkovContinuous}
\begin{equation} \label{eq:complete_fp_maintext}
    \rho_{\text{complete}} = \frac{1}{1 + \sum^{N-1}_{j=1} \Pi^{j}_{k=1} r_k}
\end{equation}
\begin{equation} \label{eq:r_halfsimplified_maintext}
    r_k = \frac{s \beta_0 \frac{1}{N} \frac{1}{2} + \delta \beta_0 \frac{1}{(N-k)\beta_0 + (k-1)\beta_1}} {s \beta_1 \frac{1}{N} \frac{1}{2} + \delta \beta_1 \frac{1}{(N-k-1)\beta_0 + k \beta_1}},
\end{equation}
for complete graphs \cite{EGTMarkovContinuous} and
\begin{equation} \label{eq:star_fp_maintext} 
    \rho_{\text{star}} = \frac{a + b}{1 + \sum^{N-2}_{j=1} c(j) \Pi^{j}_{k=1} d(j)},
\end{equation}
for star graphs \cite{EGTMarkovContinuous}, where
\begin{equation} \label{eq:a_unsimplified_maintext2}
    a = \frac{1}{N} \frac{ \frac{s}{2} \frac{1}{N} \beta_1 + \delta}{ \frac{s}{2} (\frac{1}{N} \beta_1 + \frac{1}{2} \beta_0) + \delta \frac{N}{N-1}}
\end{equation}
\begin{equation}  \label{eq:b_unsimplified_maintext2}
    b = \frac{N}{N-1} \frac{\frac{s}{2} \frac{1}{2} \beta_1 + \delta \beta_1 \frac{1}{(N-2)\beta_0 + \beta_1}}{ \frac{s}{2} (\frac{1}{2}\beta_1 + \frac{1}{N} \beta_0) + \delta \frac{(N-2)\beta_0 + 2 \beta_1}{(N-2)\beta_0 + \beta_1} }
\end{equation}
\begin{equation}  \label{eq:c_unsimplified_maintext2}
    c(j) = \frac{\frac{s}{2} \frac{1}{2} \beta_0 + \delta \beta_0 \frac{1}{(N-1-j)\beta_0 + j\beta_1}}{\frac{s}{2} (\frac{1}{2} \beta_0 + \frac{1}{N} \beta_1) + \delta \frac{(N-j)\beta_0 + j \beta_1}{(N-1-j) \beta_0 + j\beta_1}}
\end{equation}
\begin{equation}  \label{eq:d_unsimplified_maintext2}
    d(k) = \frac{\frac{s}{2} \frac{1}{N} \beta_0 + \delta}{ \frac{s}{2} \frac{1}{N} \beta_1 + \delta} \frac{\frac{s}{2} (\frac{1}{N} \beta_1 + \frac{1}{2} \beta_0) + \delta \frac{(N-k)\beta_0 + k\beta_1}{(N-1-k)\beta_0 + k\beta_1}}{\frac{s}{2} (\frac{1}{N} \beta_0 + \frac{1}{2} \beta_1) + \delta \frac{(N-1-k)\beta_0 + (k+1)\beta_1}{(N-1-k)\beta_0 + k\beta_1}}.
\end{equation}
\newline \newline 
Using this continuous-time EGT model, we will investigate what causes graphs to change from amplifiers to suppressors as the update rule changes from Bd to dB, by linking amplification/suppressing affects to the underlying biological parameters of birth, death, and competition.

\section{Results} \label{results:discrete_time_EGT}
\subsection{The problem: Different update rules influence whether stars amplify or suppress selection} \label{results:the_problem_discrete_model}
\begin{figure}
    \centering
    \includegraphics[scale=0.6]{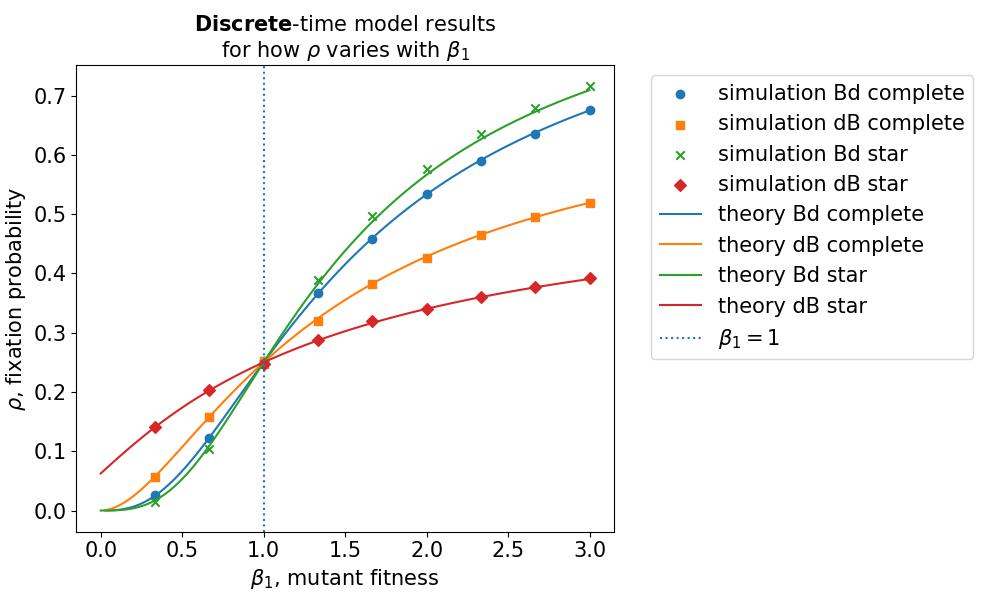}
    \caption{How the fixation probability of a mutant $\rho$ changes by varying mutant's fitness $\beta_1$ for two different update rules, Bd (birth-death with fitness on birth) and dB (death-birth with fitness on birth). The fitness of the resident was set as $\beta_0=1$. The number of nodes is $N=4$.  The numerical results of fixation probability were calculated from 100000 simulations and are compared to the known analytical solutions (complete and star graph with self-loops from  \cite{RandomGraphsBDDynamics,EGTMarkovContinuous}).}
    \label{fig:discrete_star_complete_BddBupdate}
\end{figure}
It is well known in the literature that changing the update rule in the discrete-time EGT model can impact the fixation probability \cite{pattni2015evolutionary,RandomGraphsBDDynamics,ManyUpdatesDirectedUndirected,StarUpdatesQualitativelyExplained}. 
\figurename\textbf{ }\ref{fig:discrete_star_complete_BddBupdate} shows how fixation probability $\rho$ varies with the fitness of the mutant $\beta_1$, for both numerical and analytical results for the Bd and dB dynamics for stars and complete graphs of size $N=4$. It can be seen that the star fits the definition of the amplifier for Bd but a suppressor for dB. 
\newline \newline 

\subsection{Continuum model from Bd to dB} \label{results:continuum}

\begin{figure}[h!]
    \centering
    \includegraphics[scale=0.4]{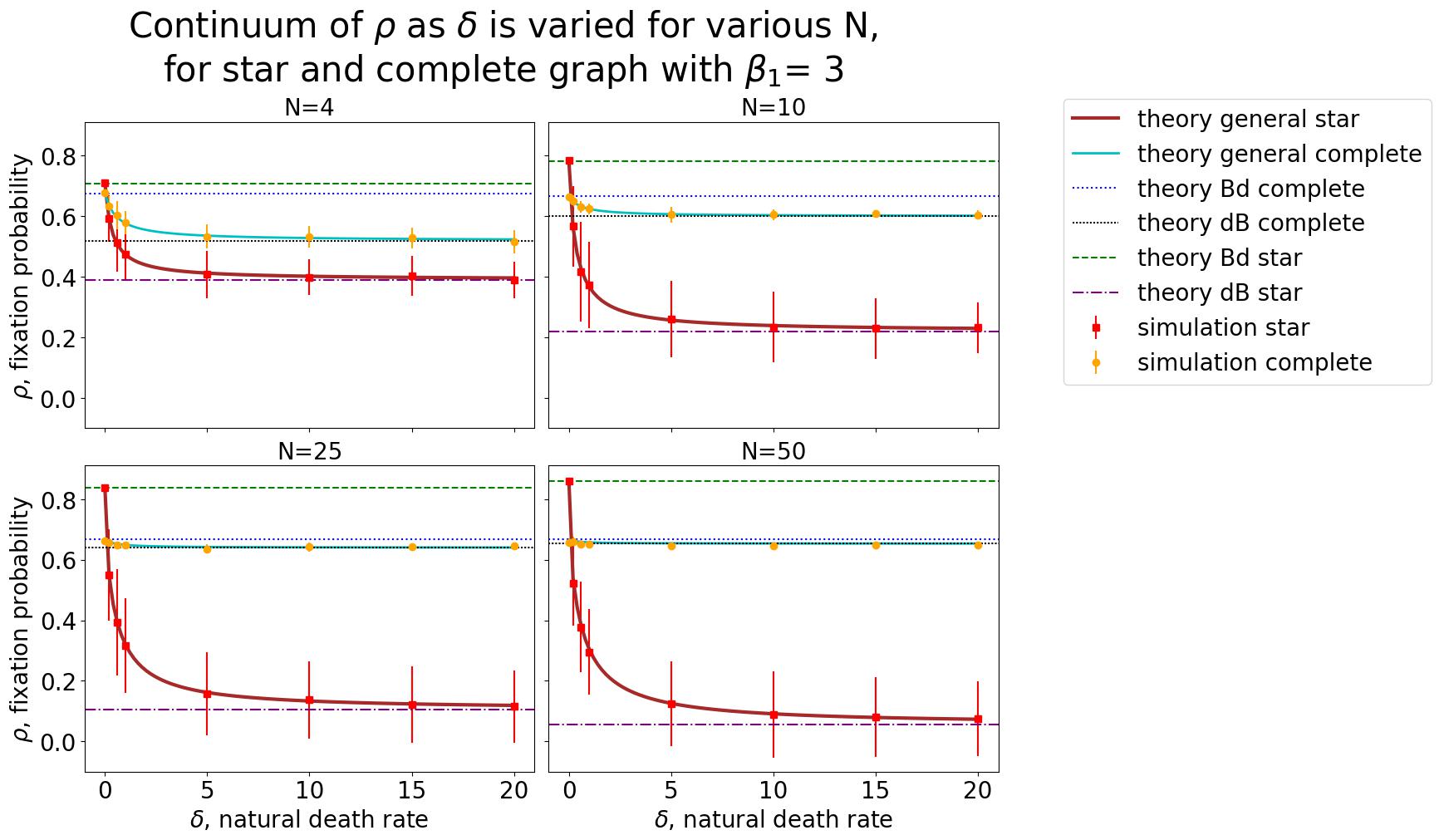}
    \caption{Showing how fixation probability $\rho$ varies with the natural death rate $\delta$ for complete and star graph. The fitness of the mutant is $\beta_1=3$, offspring survival rate is  $s=1$, number of nodes $N=4, 10, 25, \text{and }40$, competition rates $\gamma=5$, amplifying rate $c=1,000,000$. The scatter points represent the mean of simulated results of $10^4$ simulations with the error bars being the standard deviation. The solid lines representing the general star and complete graph equations that were defined by equation \ref{eq:star_fp_maintext} and  \ref{eq:complete_fp_maintext} in section \ref{sec:continuous_time_model}. The horizontal lines are the analytical solutions of Bd and dB dynamics of the discrete-time EGT model. The errors represent standard deviation (method of calculation described in section \ref{appendix:uncertainty_graphs}).
    }
    \label{fig:contiuumBddB}
\end{figure}
We want to understand why a star is an amplifier for Bd and a suppressor for dB. Since the continuous-time EGT model yields Bd and dB as special cases, we will use it to investigate the biological drivers behind this qualitative change. Setting $\delta=0$ and $s=1$ or $\delta=1$ and $s=0$ has been shown to yield Bd or dB dynamics, respectively \cite{EGTMarkovContinuous}.  However, these are strict assumptions. What if we assume that offspring always survive, taking $s=1$, then the replacement dynamics becomes 
\begin{equation} \label{eq:replacement_rate_sim_2_maintext}
    r(u,v,n_i,n_j) = \left( \beta_{u} w_{n_i,n_j} \frac{1}{2} \right) + \delta \left( \frac{\beta_{u} w_{n_i,n_j}}{\sum_{0}^{k \setminus \{j\}} \beta_{k} w_{n_k,n_j}} \right). 
\end{equation}
The first term governs the Bd dynamics and the second term governs dB dynamics. Taking $\delta=0$ results in Bd dynamics \cite{EGTMarkovContinuous}.  
If $\delta \longrightarrow \infty$ then the second term is going to be much larger then the first term, and then the replacement rate tends to (section \ref{appendix:continuum_model}) 
\begin{equation} \label{eq:replacement_rate_sim_3_maintext}
    r(u,v,n_i,n_j) \longrightarrow  \delta \left( \frac{\beta_{u} w_{n_i,n_j}}{\sum_{0}^{k \setminus \{j\}} \beta_{k} w_{n_k,n_j}} \right),
\end{equation}
which means that as $\delta \longrightarrow \infty$ we obtain dB dynamics. Therefore, there exists a continuum between Bd and dB dynamics with those being the limiting cases for $\delta=0$ and $\delta = \infty$. This is true for all graphs. 
\newline \newline 
Since increasing the intrinsic mortality rate (natural death rate $\delta$) changes the dynamics from Bd to dB, this parameter drives the star from an amplifier to a suppressor.  
This has been verified in numerical simulations seen in \figurename\textbf{ }\ref{fig:contiuumBddB} taking the star graph as an example. At $\delta=0$, the results are the same as Bd dynamics. As $\delta$ is increased from zero, $\rho$ approaches dB dynamics eventually converging to dB when $\delta$ is large enough.

\subsection{Existence of a critical mortality rate} \label{results:critical_delta}
\begin{figure}[h!]
    \centering
    \includegraphics[scale=0.4]{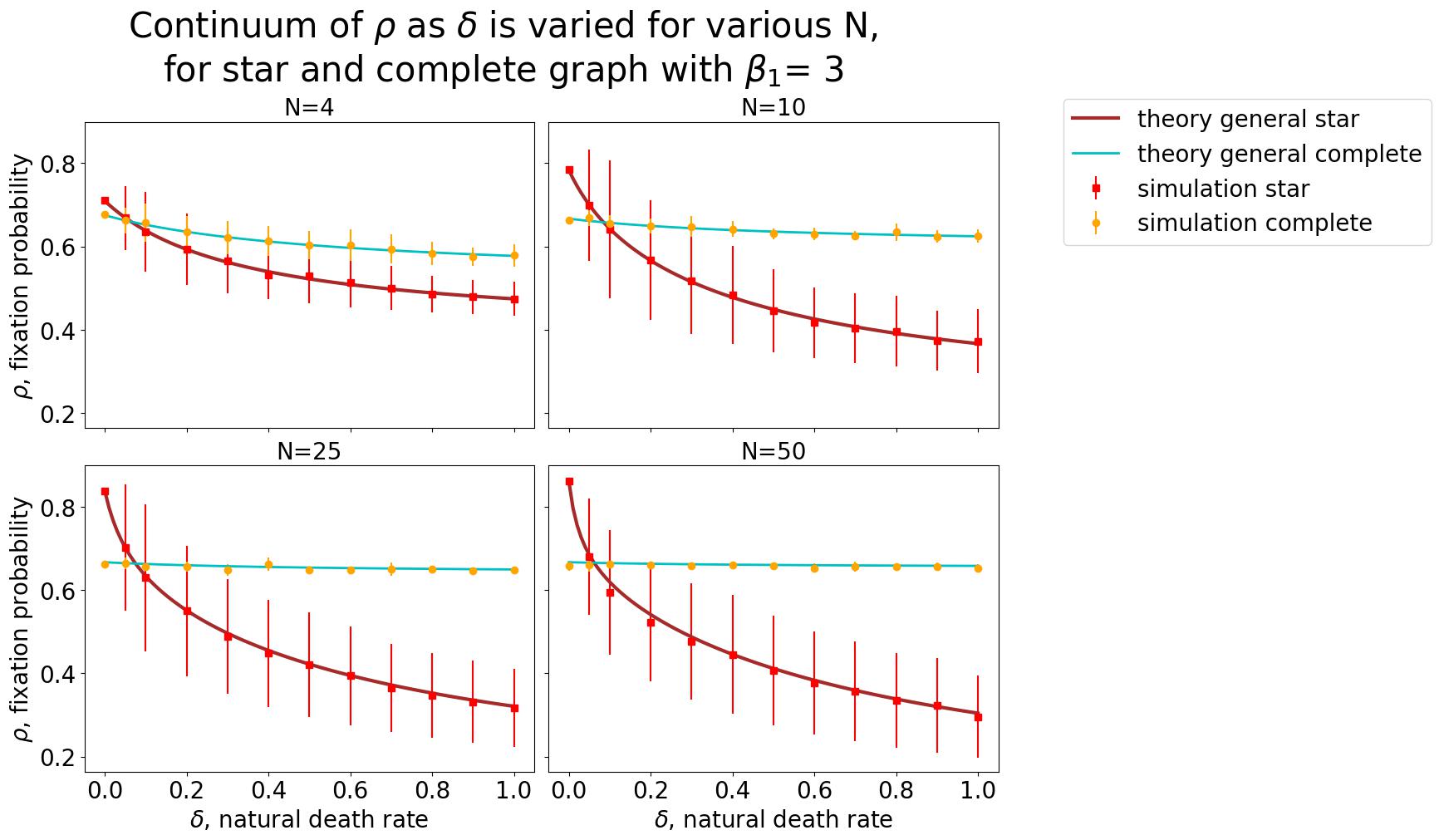}
    \caption{This shows the same results as in \figurename\textbf{ }\ref{fig:contiuumBddB} but focusing on range between of $\delta$ from 0 to 1 so where star and complete graph lines cross is more clearly seen. Where the two lines cross is where the change from amplifier to suppressor occurs, and happens at a critical value of $\delta=\delta_c$
    As in \figurename\textbf{ }\ref{fig:contiuumBddB}, the fitness of the mutant is $\beta_1=3$, offspring survival rate is  $s=1$, number of nodes $N=4$, competition rates $\gamma=5$, amplifying rate $c=1,000,000$. The scatter points represent the mean of simulated results of $10^4$ simulations with the error bars being the standard deviation. The solid lines representing the general star and complete graph equations that were defined by equation \ref{eq:star_fp_maintext} and  \ref{eq:complete_fp_maintext} in section \ref{sec:continuous_time_model}. The analytical solutions of Bd and dB dynamics of the discrete-time EGT model were emitted for clarity. The errors represent standard deviation (method of calculation described in section \ref{appendix:uncertainty_graphs}) }
    \label{fig:contiuumBddB_zoomzerotoone}
\end{figure}

\figurename\textbf{ }\ref{fig:contiuumBddB_zoomzerotoone} shows the same results as \figurename\textbf{ }\ref{fig:contiuumBddB} but focusing on small $\delta$ range from 0 to 1. In \figurename\textbf{ }\ref{fig:contiuumBddB_zoomzerotoone} it can be observed that the fixation probabilities for complete graphs ($\rho_{\text{complete}}$) and star graphs($\rho_{\text{star}}$) intersect. For low values of $\delta$, the star is an amplifier (since $\rho_{\text{star}} > \rho_{\text{complete}}$), but for large values of $\delta$ star is a suppressor (since $\rho_{\text{star}} < \rho_{\text{complete}}$). The change from an amplifier to suppressor occurs when the fixation probabilities are equal ($\rho_{\text{star}} = \rho_{\text{complete}}$). The value of $\delta$ where this occurs will be referred to to as the critical mortality rate, which we denote $\delta_c$. Since both fixation probabilities are continuous functions of $\delta$ (Equations~\eqref{eq:complete_fp_maintext} and~\eqref{eq:star_fp_maintext}), and we start with $\rho_{\text{star}} > \rho_{\text{complete}}$ under Bd dynamics and end with $\rho_{\text{star}} < \rho_{\text{complete}}$ under dB dynamics, this critical mortality rate will exist.

\subsection{Finding an upper-bound for the critical mortality rate} \label{results:critical_delta_upper_bound}
To understand the magnitude of the critical mortality rate, here we obtain an upper bound on $\delta_c$ for large populations. To determine the amplification/suppression affect, we consider the difference between the fixation probabilities of a mutant for star (equation \ref{eq:star_fp_maintext}, section \ref{sec:continuous_time_model}) and complete graphs (equation \ref{eq:complete_fp_maintext}, section \ref{sec:continuous_time_model}). This is given by \cite{EGTMarkovContinuous} 
\begin{equation} \label{eq:star_complete_difference_maintext}
\begin{aligned} 
    \rho_{\text{star}} - \rho_{\text{complete}} = \\ \frac{ (a+b - 1) + (a+b)\sum^{N-1}_{j=1} \Pi^{j}_{k=1} r_k - \sum^{N-2}_{j=1} c(j) \Pi^{j}_{k=1} d(k)}{1 + \sum^{N-2}_{j=1} c(j) \Pi^{j}_{k=1} d(k) + \sum^{N-1}_{j=1} \Pi^{j}_{k=1} r_k + (\sum^{N-2}_{j=1} c(j) \Pi^{j}_{k=1} d(k))(\sum^{N-1}_{j=1} \Pi^{j}_{k=1} r_k)} \text{,}
\end{aligned}
\end{equation}
where
\begin{equation} \label{eq:a_unsimplified_maintext}
    a = \frac{1}{N} \frac{ \frac{s}{2}  \frac{1}{N} \beta_1 + \delta}{ \frac{s}{2} ( \frac{1}{N} \beta_1 + \frac{1}{2} \beta_0) + \delta \frac{N}{N-1}}\text{,}
\end{equation}
\begin{equation}  \label{eq:b_unsimplified_maintext}
    b = \frac{N}{N-1} \frac{\frac{s}{2} \frac{1}{2} \beta_1 + \delta \beta_1 \frac{1}{(N-2)\beta_0 + \beta_1}}{ \frac{s}{2} (\frac{1}{2} \beta_1 +  \frac{1}{N} \beta_0) + \delta \frac{(N-2)\beta_0 + 2 \beta_1}{(N-2)\beta_0 + \beta_1} }\text{,}
\end{equation}
\begin{equation}  \label{eq:c_unsimplified_maintext}
    c(j) = \frac{\frac{s}{2} \frac{1}{2} \beta_0 + \delta \beta_0 \frac{1}{(N-1-j)\beta_0 + j\beta_1}}{\frac{s}{2} (\frac{1}{2} \beta_0 +  \frac{1}{N} \beta_1) + \delta \frac{(N-j)\beta_0 + j \beta_1}{(N-1-j) \beta_0 + j\beta_1}}\text{,}
\end{equation}
\begin{equation}  \label{eq:d_unsimplified_maintext}
    d(k) = \frac{\frac{s}{2}  \frac{1}{N} \beta_0 + \delta}{ \frac{s}{2}  \frac{1}{N} \beta_1 + \delta} \frac{\frac{s}{2} ( \frac{1}{N} \beta_1 + \frac{1}{2} \beta_0) + \delta \frac{(N-k)\beta_0 + k\beta_1}{(N-1-k)\beta_0 + k\beta_1}}{\frac{s}{2} ( \frac{1}{N} \beta_0 + \frac{1}{2} \beta_1) + \delta \frac{(N-1-k)\beta_0 + (k+1)\beta_1}{(N-1-k)\beta_0 + k\beta_1}}\text{,}
\end{equation}
\begin{equation} \label{eq:r_halfsimplified_maintext2}
    r_k = \frac{s \beta_0 \frac{1}{N} \frac{1}{2} + \delta \beta_0 \frac{1}{(N-k)\beta_0 + (k-1)\beta_1}} {s \beta_1 \frac{1}{N} \frac{1}{2} + \delta \beta_1 \frac{1}{(N-k-1)\beta_0 + k \beta_1}}\text{.}
\end{equation}
Note that we assume $\delta$ is the same for all individuals types \cite{EGTMarkovContinuous}, so fitness only depends on the birth rates ($\beta_0$ and $\beta_1$).
\newline \newline 
We will consider the large population size limit, i.e.  $N \longrightarrow \infty$. The fitness coefficients are taken as $\beta_0=1$, $\beta_1=r$  \cite{RandomGraphsBDDynamics,DynamicsStructuredPopulations,ReviewGraphGameTheoryApplication}.
As discussed in section \ref{results:continuum}, we will take $s=1$ so the offspring survives regardless of the number of pre-existing individuals occupying the node it is placed on \cite{EGTMarkovContinuous}. Then equation  \ref{eq:star_complete_difference_maintext} simplifies into the following (section \ref{appendix:upper_bound_critical})
\begin{equation} \label{eq:final_difference_equation_maintext}
   \rho_{\text{star}} - \rho_{\text{complete}} = -\frac{4\delta(r-1)}{r(r+4\delta)} \text{.}
\end{equation}
This relationship between $\rho_{\text{star}}$ and $\rho_{\text{complete}}$ only holds for
\begin{equation} \label{eq:true_hold_maintext}
  \delta \gg \frac{1}{\sqrt{N}} \text{.}
\end{equation}
We assume an advantageous mutant, so $r>0$. For $\delta=0$, we have Bd dynamics with $\rho_{\text{star}} - \rho_{\text{complete}}>0$, so the star is an amplifier. For $\delta \gg \frac{1}{\sqrt{N}}$, $\rho_{\text{star}} - \rho_{\text{complete}}<0$, so the star is a suppressor. Since for all $\delta$ satisfying equation~\ref{eq:true_hold_maintext} the star is a suppressor of selection, this gives an upper bound on $\delta_c$, i.e. $\delta_c < \frac{1}{\sqrt{N}}$. This means for sufficiently large populations with $ \frac{1}{\sqrt{N}} \approx 0$, then the critical value $\delta_c \longrightarrow 0$ and the star is a suppressor for any $\delta > 0$.   

\subsection{Numerical simulations} \label{results:numerical_final}

\begin{figure}
    \centering
    \includegraphics[scale=0.4]{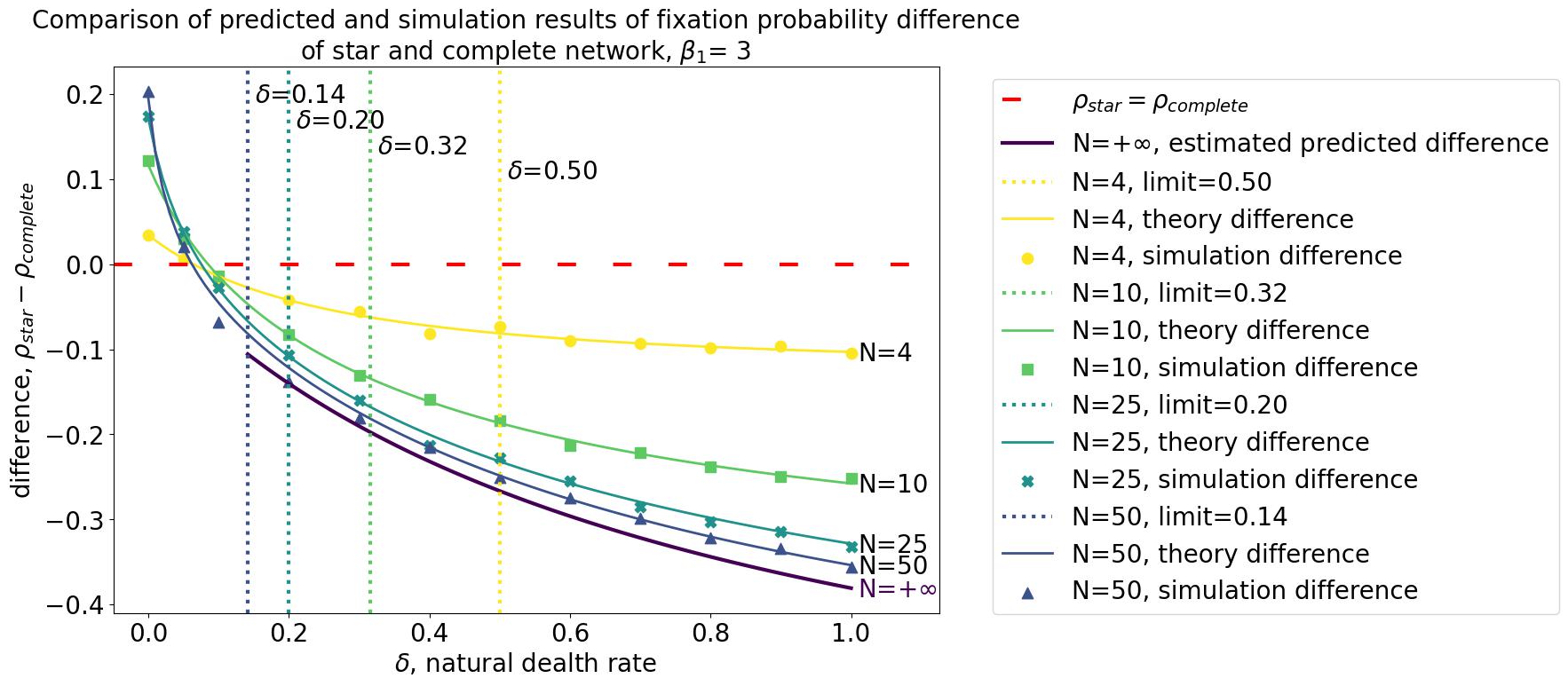}
    \caption{Showing how the difference of fixation probability of star and complete graph, $\rho_{\text{star}} - \rho_{\text{complete}}$, varies with the natural death rate $\delta$ between $0$ and $1$. The number of nodes $N=4,10,25,50$ are each represented by a specific colour. The wider red line (estimated predicted difference) represents the prediction of equation \ref{eq:final_difference_equation_maintext} (section \ref{results:critical_delta_upper_bound}) which is true for a graph of infinite size. The scatter points represent the numerical fixation probability of $10^4$ simulations, the same as \figurename\textbf{ }\ref{fig:contiuumBddB_zoomzerotoone}. The solid lines of respective colours represent the analytical results ($\rho_{\text{star}} - \rho_{\text{complete}}$, with $\rho_{\text{star}}$ defined by equation \ref{eq:star_fp_maintext} and $\rho_{\text{complete}}$ defined by equation \ref{eq:complete_fp_maintext}, section \ref{sec:continuous_time_model}) for that size of graph (number of nodes). The vertical dotted lines represent the upper bound limit of the change from amplifier to suppressor, as defined by equation \ref{eq:true_hold_maintext} and is dependent on $N$. 
    The dashed green line represents the point of $\rho_{\text{star}} - \rho_{\text{complete}}=0$, which is important to note as this where the change from amplifier to suppressor occurs. 
    The parameter values are: the fitness of mutant $\beta_1=3$, offspring survival $s=1$, competition rate $\gamma=5$, amplifying rate $c=1000000$, the graph coefficients $w=w_c = \frac{1}{N}$ and $w_l=\frac{1}{2}$, reruns $10^4$. Error bars were removed for clarity and an error bar version is seen in section \ref{appendix:additional_graphs}.
     }
    \label{fig:fixation_prob_difference_theory_sim}
\end{figure}

Here we investigate the impact of $\delta$ on whether a star graph amplifies or suppresses selection numerically for finite population sizes. 
\newline \newline 
The difference between the fixation probabilities of star and complete graphs ($\rho_{\text{star}}-\rho_{\text{complete}}$) of size $N=4,10,25,50$ is plotted for values of $\delta$ between $0$ and $1$ in \figurename\textbf{ }\ref{fig:fixation_prob_difference_theory_sim}. It is compared to equation \ref{eq:final_difference_equation_maintext} (section \ref{results:critical_delta_upper_bound}) represented by the wide red line, which is true for an infinite size graph ($N=\infty$). It can be observed that as $N$ increases, the difference $\rho_{\text{star}}-\rho_{\text{complete}}$ gets closer and closer to the predicted difference of equation \ref{eq:final_difference_equation_maintext}. Hence for a large graph, equation \ref{eq:final_difference_equation_maintext} would be able to describe the system.
\newline \newline
We note that the star changes from amplifier to suppressor somewhere between $\delta=0$ and the limit $\delta=\frac{1}{\sqrt{N}}$, which is the upper bound on where the change would happen that was predicted as $N \to \infty$. Therefore, this appears to provide a robust upper bound to the critical value, $\delta_c$. As the graph size increases, the upper bound shifts to the left towards $0$, demonstrating that for larger population sizes, smaller death rates are required to drive the star from an amplifier to a suppressor.

\newpage

\section{Discussion}
One of the problems of the classical discrete-time EGT model is that the update rules can have a significant impact on whether graphs amplify or suppress natural selection \cite{RandomGraphsBDDynamics, ManyUpdatesDirectedUndirected,StarUpdatesQualitativelyExplained,yagoobi2023categorizing,allen2020transient}. When investigating how the structure impacts the spread of mutation, having a model that differs significantly depending on the order of birth and death events (the update rule) is a problem, especially if wanting to use these models to predict real biological phenomena. If investigating a real biological population that exhibited a star-graph-like structure, the discrete-time EGT model would not be able to answer whether the structure would be an amplifier or suppressor of mutation unless the update rule of the system is known. However, experimentally an update rule might not be possible to measure, since it does not depend on measurable biological quantities.
\newline \newline 
This problem has been extensively studied in the literature, including understanding why a star graph behaves differently depending on the update rule \cite{RandomGraphsBDDynamics, ManyUpdatesDirectedUndirected,StarUpdatesQualitativelyExplained}. Other graph types have also been studied looking at the Birth-Death with fitness on birth (Bd) and Death-Birth with fitness on birth (dB) update rules. It has been found that small (graphs of size $N \leq 14$) undirected (unweighted) random graphs are generally amplifiers under Bd and suppressors under dB \cite{RandomGraphsBDDynamics,bhaumik2024constant}. Alcalde  et al. (2018)  \cite{alcalde2018evolutionary} is an extensive study looking at a database (found at Alcalde  et al. (2017) \cite{alcalde2017accurate}) of undirected graphs of size $N \leq 10$ under Bd. This is a total of 11,989,763 graphs excluding the graph of one vertex, finding that most of these graphs are amplifiers. A small portion of graphs were found to transition from amplifier to suppressor (and vice versa) at a critical value of mutant fitness, with some graphs having multiple transitions. For example, from 11,117 $N=8$ graphs, 10,544 (94.85\%) were found to be amplifiers, a large proportion. The rest, 573 graphs (5.15\%), were of other types, with 466 graphs (4.19\%) having one or more transitions. The majority of those change from suppressor to amplifier (427 graphs, 3.84\%), and a smaller portion of the opposite change from amplifier to suppressor (36 graphs, 0.32\%). Only 3 graphs (0.03\%) showed more than one transition of changing from suppressor to amplifier back to suppressor. It is also interesting to note that transitions did not appear until reaching graph size of six nodes \cite{alcalde2018evolutionary}. 
\newline \newline 
The size of the population has also been found to play a role in whether graphs are likely to amplify or suppress selection. Although random graphs have been found to be amplifiers under Bd for small values of $N$, Adlam and Nowak (2014) \cite{adlam2014universality} found that increasing $N$ leads to the fixation probability approaching that of homogeneous population for random graphs (Erdős–Rényi model, shown numerically and analytically), random graphs with small-world properties (Watts-Strogatz model, shown numerically) and scale-free graphs (Barabási–Albert model, shown numerically).  It is interesting to note that the difference between Bd and dB for star graphs does not diminish for large $N$ \cite{StarUpdatesQualitativelyExplained}. 
 \newline \newline 
 Different modifications to the model that result in suppression have been studied. Other update rules have also been studied with additional modifications, such as fitness being linked to the death event \cite{pattni2015evolutionary,ManyUpdatesDirectedUndirected}, fitness affecting both birth and death events \cite{kaveh2015duality}, and migration of the parent instead of the offspring during the birth event \cite{sharma2025graph}. By considering six types of update rules, it has been found for star graphs that the amplification is strongest under Bd and the suppression is strongest under dB \cite{ManyUpdatesDirectedUndirected,StarUpdatesQualitativelyExplained}. 
The use of a particular update rule at times has been justified by an assumption based on specific behaviour of real biological systems. For example, having the update rule where parental migration occurs instead of offspring related to behaviour of snakes and lizards abandoning their eggs \cite{sharma2025graph}. Due to the difference between cancerous and non-cancerous cells, having a model with fitness impacting both birth and death rates was considered more suitable \cite{kaveh2015duality}. 
\newline \newline 
Another factor that can impact whether a graph is classed an amplifier or suppressor is the method for deciding a mutant's initial placement on the graph. It has been found that for temperature initialisation (where initial location is proportional to the degree of each node) a star graph is a suppressor of selection \cite{adlam2015Amplifiersofselection}. Models that decouple migration from birth or death events have found that asymmetric \cite{marrec2021toward} and large migration rates \cite{abbara2023frequent} can have suppressing effects on how mutants spread.  
\newline \newline 
Using a continuous-time EGT model, which has been proven to be able to recreate results of Bd and dB update rule of classical discrete-time EGT model~\cite{EGTMarkovContinuous}, we have shown that there exists a continuum between the Bd and dB update rules, which is controlled by the natural death rate. When taking the natural death rate as equal to zero, we recover Bd dynamics, and as the natural death rate tends to infinity, we have shown that the system recovers dB dynamics. This suggests that it is the natural death rate that drives suppression of selection in graphs. Using the star graph as an example, we have shown this numerically. There is a change from an amplifier to a suppressor as the natural death rate is increased, which is shown by the fixation probability of the star changing from being larger than that of the complete graph to smaller. This suggests that there must be a critical natural death rate where this transition occurs from amplification to suppression. 
\newline \newline 
By taking the large population size limit, we have derived an upper bound for the critical natural death rate, equal to $\frac{1}{\sqrt{N}}$, where $N$ is the population size. Above this, the star graph will always suppress selection. Based on this upper bound, if the population size $N$ is very large, the critical natural death rate must be negligible. Since in most biological systems the natural death rate is likely to be greater than zero, this implies that the star is likely to be a suppressor for large populations.  
\newline \newline
This work has only focused on star graphs as an example of an amplifier under Bd and suppressor under dB. It would be interesting to see our results in the context of other graphs, such as undirected random graphs of various sizes, graphs that are amplifiers for both Bd and dB update rules, as found here \cite{svoboda2024amplifiers}, or graphs that are amplifiers in dB but not Bd \cite{svoboda2024amplifiers}. This work has also only considered the continuous-time EGT model under assumptions that are compatible with the classical discrete-time EGT model \cite{EvolutionaryDynamicsGraphs}. It would be interesting to consider our results under other model assumptions, such as decoupling of migration rates, considering more than one individual existing on the nodes (effectively increasing carrying capacity), temperature initialization of initial mutant placement, and directed graphs. Considering these things would remove some of the limitations of the model, such as having one individual per node and migration being coupled to birth, which might also make the model more applicable in realistic biological applications. 

\section{Conclusion}
In evolutionary graph theory, it is well known that the choice of update rate (i.e. the order in which birth and death events take place) can lead to qualitatively different conclusions on the impact of population structure on the evolutionary process. For example, under birth-death with fitness of birth (Bd), the star graph is an amplifier of natural selection, but under death-birth with fitness on birth (dB), the star graph is a suppressor of natural selection. In this paper, we have found that there exists a biologically motivated continuum model between the Bd and dB update rules in evolutionary graph theory. We showed that the natural death rate drives the transition from Bd to dB dynamics. Using the example of a star graph, we have shown that for large population sizes, a star graph would be a suppressor of natural selection unless the natural death rate is negligible. 

\section*{Acknowledgments and Funding}
MD acknowledgments funding by Faculty funded PGR studentship given by Department of Mathematics, School of Physical Sciences, University of Liverpool. \\
The authors would like to thank the use of HTCondor Pool that forms part of the Research IT facilities at the University of Liverpool, UK (\url{https://www.liverpool.ac.uk/research-it/high-throughput-computing/}), which significantly sped up simulation of the stochastic models.

\bibliography{bibliography}

\section*{Appendices}
\begin{appendices}
\section{Proof of continuum model} \label{appendix:continuum_model}
When the amplifier $c \longrightarrow 0$, the replacement rate $r$ of an individual to become another individual, let's say for a resident $u$ on node $n_i$ to be replaced by a mutant $v$ whose parent is on $n_j$, is defined as \cite{EGTMarkovContinuous}
\begin{equation} \label{eq:replacement_rate}
\begin{split}
r & = r(u,v,n_i,n_j) \\
r(u,v,n_i,n_j) & = s \left( \beta_{u} w_{n_i,n_j} \frac{\delta_{v} + \gamma_{v,u}}{\delta_{v} + \gamma_{v,u} + \delta_u + \gamma_{u,v}} \right) + \delta_{v} \left( \frac{\beta_{u} w_{n_i,n_j}}{\sum_{0}^{k \setminus \{j\}} \beta_{k} w_{n_k,n_j}} \right)  \\
r(u,v,n_i,n_j)& = A \times \frac{B}{C} + \text{D} \times \frac{\text{E}}{F}  
\end{split}
\end{equation}
where 
\begin{itemize}
    \item A is the birth rate for $u$ on node $n_i$ with offspring to be placed and surviving on $n_j$ 
    \item B is the death rate of $v$
    \item C is the total death rate of $u$ and $v$
    \item D is the natural death rate of an individual of $v$
    \item E is product of fitness of $v$ and the graph weight edge between $n_i$ and $n_j$
    \item F is sum of product of birth rates and graph rates excluding $u_j$ 
\end{itemize}
In EGT we require replacement probability rather than rates. To get the probability of replacement of individual $u$ by individual $v$ we would need to divide by the sum of all possible replacements rates, hence the probability of replacement is defined as \cite{EGTMarkovContinuous} 
\begin{equation} \label{eq:def_replacement_probability}
    R(u,v) = \frac{r(u,v,n_i,n_j)}{\sum_{f \in S} \sum_{k \in S}  r(f,k,n_f,n_k)}
\end{equation}
where $S$ is the set of all the individuals' traits.\cite{EGTMarkovContinuous} 
\newline \newline 
If $\delta_i=\delta_j = \delta$ and $\gamma_{i,j} = \gamma_{j,i} = \gamma$, then
\begin{equation}
\begin{split}
&  \frac{\delta_j + \gamma_{j,i}}{\delta_j + \gamma_{j,i} + \delta_i + \gamma_{i,j}} \\
& = \frac{\delta + \gamma}{\delta + \gamma + \delta + \gamma} \\
& =  \frac{1(\delta+ \gamma)}{2(\delta+ \gamma)} \\
& = \frac{1}{2}
\end{split}
\end{equation}
Then 
\begin{equation} \label{eq:replacement_rate_sim}
    r(i,j,n_i,n_j) = s \beta_i w_{n_i,n_j} \frac{1}{2} + \frac{\delta_j \beta_i w_{n_i,n_j}}{\sum_{k \neq j}  \beta_k w_{n_k,n_j}}
\end{equation}
Taking the replacement rate \ref{eq:replacement_rate_sim} and substituting it into \ref{eq:def_replacement_probability} gives \cite{EGTMarkovContinuous} 
\begin{equation} \label{eq:replacement_probability_not_simplified}
\begin{split}
    R(i,j)& = \frac{ r(i,j,n_i,n_j)}{\sum_l \sum_m r(l,m,n_l,n_m)} \\
    R(i,j)& = \frac{ s \beta_i w_{n_i,n_j} \frac{1}{2} + \frac{\delta_j \beta_i w_{n_i,n_j}}{\sum_{k \neq j}  \beta_k w_{n_k,n_j}} }
    {\sum_l \sum_m  s \beta_l w_{n_l,n_m} \frac{1}{2} + \frac{\delta_m \beta_l w_{n_l,n_m}}{\sum_{k \neq m}  \beta_k w_{n_k,n_m}}}
\end{split}   
\end{equation}
The denominator can be rewritten as 
\begin{align*}
\begin{split}
& \sum_l \sum_m  s \beta_l w_{n_l,n_m} \frac{1}{2} + \frac{\delta_m \beta_l w_{n_l,n_m}}{\sum_{k \neq m}  \beta_k w_{n_k,n_m}} \\
& = \sum_l \sum_m \beta_l w_{n_l,n_m} s \frac{1}{2}
+ \sum_l \sum_m \delta_m \frac{\beta_l w_{n_l,n_m}}{\sum_{k \neq m}  \beta_k w_{n_k,n_m}}
\end{split}
\end{align*}
We note that an individual who has died, cannot then produce the offspring, so for the second sum of the second term we impose an extra condition on the sum \cite{EGTMarkovContinuous}
\begin{align*}
 = \sum_l \sum_m \beta_l w_{n_l,n_m} s \frac{1}{2}
+ \sum_l \sum_{m \neq n} \delta_m \frac{\beta_l w_{n_l,n_m}}{\sum_{k \neq m}  \beta_k w_{n_k,n_m}}
\end{align*}
We note that $r(l,m,n_l,n_m) = r(m,l,n_m,n_l)$\cite{EGTMarkovContinuous}, which means
\begin{align*}
\delta_m \frac{\beta_l w_{n_l,n_m}}{\sum_{k \neq m}  \beta_k w_{k,m}} = 
\delta_l \frac{\beta_m w_{n_m,n_l}}{\sum_{k \neq l} \beta_k w_{k,l}}
\end{align*}
So the denominator can be rewritten as
\begin{align*}
\sum_l \sum_m \beta_l w_{n_l,n_m} s \frac{1}{2}
+ \sum_l \sum_{m \neq l} \delta_l \frac{\beta_m w_{n_m,n_l}}{\sum_{k \neq l} \beta_k w_{n_k,n_l}}
\end{align*}
We note that in the first term $s \frac{1}{2}$ are independent of $l$ and $m$ and $\beta_l$ is independent of $m$. For the second term, $\delta_l$ and $\sum_{k \neq l} \beta_k w_{k,l}$ is independent of $m$. This means that the denominator can be rearranged as 
\begin{align*}
s \frac{1}{2} \sum_l \beta_l \sum_m  w_{n_l,n_m} 
+ \sum_l \delta_l \frac{ \sum_{m \neq l} \beta_m w_{n_m,n_l}}{\sum_{k \neq l} \beta_k w_{n_k,n_l}}
\end{align*}
As $k=m$, then $\frac{ \sum_{m \neq l} \beta_m w_{n_m,n_l}}{\sum_{k \neq l} \beta_k w_{n_k,n_l}} = 1$, so then the denominator is equal to  
\begin{align} \label{eq:replacement_probability_denominator_simplifier}
s \frac{1}{2} \sum_l \beta_l \sum_m  w_{n_l,n_m} + \sum_l \delta_l
\end{align}
If $l=N$ and $\delta_l$ is the same for all individuals then 
\begin{align} \label{eq:replacement_probability_denominator_simplifier_v2}
s \frac{1}{2} \sum_l \beta_l \sum_m  w_{n_l,n_m} + N\delta
\end{align}
Substituting \ref{eq:replacement_probability_denominator_simplifier_v2} back into \ref{eq:replacement_probability_not_simplified} gives 
\begin{align} \label{eq:replacement_probability_simplified}
    R(i,j) =  \frac{s \frac{1}{2} \beta_i w_{n_i,n_j} + \delta_j \frac{\beta_i w_{n_i,n_j}}{\sum_{k \neq j}  \beta_k w_{n_k,n_j}}}
{s \frac{1}{2} \sum_l \beta_l \sum_m  w_{n_l,n_m} + N\delta}
\end{align}
We note that the replacement probability of Bd is \cite{EGTMarkovContinuous,pattni2015evolutionary}
\begin{equation} \label{eq:replacement_probability_Bd}
    R(i,j)_{Bd} = \frac{\beta_i w_{i,j}}{\sum_n \beta_n w_{i,n}}
\end{equation}
The replacement probability of dB is \cite{EGTMarkovContinuous,pattni2015evolutionary}
\begin{equation} \label{eq:replacement_probability_dB}
    R(i,j)_{dB} = \frac{1}{N}\frac{\beta_i w_{i,j}}{\sum_n \beta_n w_{n,j}}
\end{equation}
\cite{EGTMarkovContinuous} proved that taking $s=1$ and $\delta=0$ gives Bd replacement probability (\ref{eq:replacement_probability_Bd}) and taking take $s=0$ and $\delta=1$ gives dB replacement probability (\ref{eq:replacement_probability_dB}). 
\newline \newline 
Now let's consider the scenario where neither $s$ or $\delta$ are zero. Assuming that $\delta$ is the same value for all individuals in a population of size $N$, and $s=1$, this gives the replacement probability 
\begin{align*}
R(i,j) = \frac{\frac{1}{2} \beta_i w_{n_i,n_j} + \delta \frac{\beta_i w_{n_i,n_j}}{\sum_{k \neq j}  \beta_k w_{n_k,n_j}}}
{ s \frac{1}{2} \sum_l \beta_l \sum_m  w_{n_l,n_m} + N\delta}
\end{align*}
Factorizing by $\delta$ gives
\begin{align*}
\begin{split}
R(i,j) & = \frac{\delta}{\delta} \frac{\frac{1}{2} \beta_i w_{n_i,n_j} \frac{1}{\delta} + \frac{\beta_i w_{n_i,n_j}}{\sum_{k \neq j}  \beta_k w_{n_k,n_j}}}
{\frac{1}{2} \sum_l \beta_l \sum_m  w_{n_l,n_m} \frac{1}{\delta} +  N }\\
& = \frac{\frac{1}{2} \beta_i w_{n_i,n_j} \frac{1}{\delta} + \frac{\beta_i w_{n_i,n_j}}{\sum_{k \neq j}  \beta_k w_{n_k,n_j}}}
{\frac{1}{2} \frac{1}{\delta} \sum_l \beta_l \sum_m  w_{n_l,n_m}  +  N }
\end{split}
\end{align*}
Taking $\delta \longrightarrow \infty$ so $\frac{1}{\delta} \approx 0$ gives 
\begin{align}
\begin{split}
R(i,j) & \approx \frac{ \frac{\beta_i w_{n_i,n_j}}{\sum_{k \neq j}  \beta_k w_{n_k,n_j}}}
{  N }\\ 
& \approx \frac{1}{N} \frac{\beta_i w_{n_i,n_j}}{\sum_{k \neq j}  \beta_k w_{n_k,n_j}}
\end{split}
\end{align}
which is the same as replacement rate for dB dynamics, \ref{eq:replacement_probability_dB}.
Hence for large value of $\delta$ the system would be in dB dynamics.

\section{Full derivation of the fixation probability difference and finding the upper bound critical value $\delta_c$} \label{appendix:upper_bound_critical}
Here we show the full derivation of the difference between star graph fixation probability and complete graph fixation probability. 
\newline \newline 
The fixation probability for the \textbf{complete} graph is given by \cite{EGTMarkovContinuous}
\begin{equation} \label{eq:complete_fp}
    \rho_{\text{complete}} = \frac{1}{1 + \sum^{N-1}_{j=1} \Pi^{j}_{k=1} r_k}
\end{equation}
where,
\begin{equation} \label{eq:r_unsimplified}
    r_k = \frac{s \beta_0 w \frac{\delta_1 + \gamma_{1,0}}{\delta_0 + \delta_1 + \gamma_{0,1} + \gamma_{1,0}} + \delta_1 \beta_0 \frac{1}{(N-k)\beta_0 + (k-1)\beta_1}} {s \beta_1 w \frac{\delta_0 + \gamma_{0,1}}{\delta_0 + \delta_1 + \gamma_{0,1} + \gamma_{1,0}} + \delta_0 \beta_1 \frac{1}{(N-k-1)\beta_0 + k \beta_1}}
\end{equation}
If $\delta=\delta_0=\delta_1$ and $\gamma=\gamma_{0,1}=\gamma_{1,0}$, then 
\begin{equation} \label{eq:r_halfsimplified}
    r_k = \frac{s \beta_0 w \frac{1}{2} + \delta \beta_0 \frac{1}{(N-k)\beta_0 + (k-1)\beta_1}} {s \beta_1 w \frac{1}{2} + \delta \beta_1 \frac{1}{(N-k-1)\beta_0 + k \beta_1}}
\end{equation}
Note that when this is the case, the competition coefficients disappear. 
For complete graphs with self-loops, $w=\frac{1}{N}$. 
\newline \newline 
The fixation probability for the \textbf{star} graph is given by \cite{EGTMarkovContinuous}
\begin{equation} \label{eq:star_fp}
    \rho_{\text{star}} = \frac{a + b}{1 + \sum^{N-2}_{j=1} c(j) \Pi^{j}_{k=1} d(j)}
\end{equation}
\begin{equation} \label{eq:a_unsimplified}
    a = \frac{1}{N} \frac{ \frac{s}{2} w_c \beta_1 + \delta}{ \frac{s}{2} (w_c \beta_1 + w_l \beta_0) + \delta \frac{N}{N-1}}
\end{equation}
\begin{equation}  \label{eq:b_unsimplified}
    b = \frac{N}{N-1} \frac{\frac{s}{2} w_l \beta_1 + \delta \beta_1 \frac{1}{(N-2)\beta_0 + \beta_1}}{ \frac{s}{2} (w_l \beta_1 + w_c \beta_0) + \delta \frac{(N-2)\beta_0 + 2 \beta_1}{(N-2)\beta_0 + \beta_1} }
\end{equation}
\begin{equation}  \label{eq:c_unsimplified}
    c(j) = \frac{\frac{s}{2} w_l \beta_0 + \delta \beta_0 \frac{1}{(N-1-j)\beta_0 + j\beta_1}}{\frac{s}{2} (w_l \beta_0 + w_c \beta_1) + \delta \frac{(N-j)\beta_0 + j \beta_1}{(N-1-j) \beta_0 + j\beta_1}}
\end{equation}
\begin{equation}  \label{eq:d_unsimplified}
    d(k) = \frac{\frac{s}{2} w_c \beta_0 + \delta}{ \frac{s}{2} w_c \beta_1 + \delta} \frac{\frac{s}{2} (w_c \beta_1 + w_l \beta_0) + \delta \frac{(N-k)\beta_0 + k\beta_1}{(N-1-k)\beta_0 + k\beta_1}}{\frac{s}{2} (w_c \beta_0 + w_l \beta_1) + \delta \frac{(N-1-k)\beta_0 + (k+1)\beta_1}{(N-1-k)\beta_0 + k\beta_1}}
\end{equation}
We assume that $\delta = \delta_0 = \delta_1$ and $\gamma = \gamma_{0,1} = \gamma_{1,0}$.
From definition of star, $w_l = \frac{1}{2}$ and $w_c = \frac{1}{N}$. \cite{EGTMarkovContinuous} 
\newline \newline 
Taking the difference of the fixation probability (equations \ref{eq:star_fp} and \ref{eq:complete_fp}) gives 
\begin{equation} \label{eq:star_complete_difference}
\begin{aligned} 
    \rho_{\text{star}} - \rho_{\text{complete}} = \\ \frac{ (a+b - 1) + (a+b)\sum^{N-1}_{j=1} \Pi^{j}_{k=1} r_k - \sum^{N-2}_{j=1} c(j) \Pi^{j}_{k=1} d(k)}{1 + \sum^{N-2}_{j=1} c(j) \Pi^{j}_{k=1} d(k) + \sum^{N-1}_{j=1} \Pi^{j}_{k=1} r_k + (\sum^{N-2}_{j=1} c(j) \Pi^{j}_{k=1} d(k))(\sum^{N-1}_{j=1} \Pi^{j}_{k=1} r_k)}
\end{aligned}
\end{equation}
To simplify equation \ref{eq:star_complete_difference}, let's take a large value of $N$, i.e. limit of $N$ to infinity, $N \longrightarrow \infty$.
\newline \newline 
Using this, the expressions \ref{eq:r_halfsimplified},\ref{eq:a_unsimplified},  \ref{eq:b_unsimplified},\ref{eq:c_unsimplified},\ref{eq:d_unsimplified} become  
\begin{equation} \label{eq:a_simplified}
    a = 0
\end{equation}
\begin{equation} \label{eq:b_simplified}
    b = \frac{s w_l \beta_1}{ s w_l \beta_1 + s w_c \beta_0 +  2\delta }
\end{equation}
\begin{equation} \label{eq:c_simplified}
    c(j) = \frac{s w_l \beta_0}{s w_l \beta_0 + s w_c \beta_1 + 2\delta}
\end{equation}
\begin{equation} \label{eq:d_simplified}
    d(k) = \left( \frac{s w_c \beta_0 + 2\delta}{ s w_c \beta_1 + 2\delta} \right) \left( \frac{s w_c \beta_1 + s w_l \beta_0 + 2\delta}{s w_c \beta_0 + s w_l \beta_1 + 2\delta} \right)
\end{equation}
\begin{equation} \label{eq:r_simplified}
    r_k = \frac{\beta_0}{\beta_1}
\end{equation}
Note that now $c$ is independent of $j$ and $d$ is independent of $k$. Therefore
\begin{equation*}
    \sum^{\infty}_{j=1} c(j) \Pi^{j}_{k=1} d(k) =  \sum^{\infty}_{j=1} c d^j  = \frac{cd}{1-d}
\end{equation*}
using the rules for geometric series, $\sum^{\infty}_{k=m} a r^k = \frac{a r^m}{1-r}$, with $|r| < 1$. \\
Similarly,
\begin{equation*}
    \sum^{\infty}_{j=1} \Pi^{j}_{k=1} r_k = \sum^{\infty}_{j=1} r_k^j = \frac{r_k}{1-r_k}
\end{equation*}
So the equation \ref{eq:star_complete_difference} becomes 
\begin{equation} \label{eq:star_complete_difference_simplified}
    \rho_{\text{star}} - \rho_{\text{complete}} = \frac{ (b - 1) + b \frac{r_k}{1-r_k} - \frac{cd}{1-d}}{1 +  \frac{cd}{1-d} +  \frac{r_k}{1-r_k} + ( \frac{cd}{1-d})( \frac{r_k}{1-r_k})}
\end{equation}
It can be noted that equation \ref{eq:d_simplified} has a $\delta$ and $\delta^2$ if the fractions are expanded and combined. 
\begin{equation*}
    d =  \frac{s w_c \beta_0 s w_c \beta_1 + s w_c \beta_0 s w_l \beta_0 + 2\delta(s w_c \beta_0 + s w_c \beta_1 + s w_l \beta_0) + 4\delta^2}{ s w_c \beta_1 s w_c \beta_0 +s w_c \beta_1 s w_l \beta_1 + 2\delta (s w_c \beta_1 +  s w_c \beta_0 + s w_l \beta_1) + 4\delta^2} 
\end{equation*}
The assumption is that $N$ is very large, so $\frac{1}{N} \approx 0$. Therefore we must assume all other variables are sufficiently larger than $\frac{1}{N}$. Since $\frac{1}{N} < 1$, a sufficient condition for $\delta$ and $\delta^2$ to be larger enough is $\delta^2 \gg \frac{1}{N}$. Hence the limiting factor is that 
\begin{equation} \label{eq:true_hold}
  \delta > \frac{1}{\sqrt{N}}
\end{equation}
\newline \newline 
As working in the framework of the continuum model, so setting $s=1$. We assume that $\beta_0=1$ and $\beta_1=r$ \cite{ReviewGraphGameTheoryApplication}. $w_c=w=\frac{1}{N}$, $w_l = \frac{1}{2}$ from graph properties \cite{EGTMarkovContinuous}. This gives
\begin{equation}
    b = \frac{r}{ r+  4\delta }
\end{equation}
\begin{equation}
    c = \frac{1}{1 + 4\delta}
\end{equation}
\begin{equation}
    d = \frac{1 + 4\delta}{r + 4\delta}
\end{equation}
\begin{equation}
    r_k = \frac{1}{r}
\end{equation}
Inputting these into equation \ref{eq:star_complete_difference_simplified} gives 
\begin{equation*}
    \rho_{\text{star}} - \rho_{\text{complete}} = \frac{ (\frac{r}{ r+  4\delta } - 1) + \frac{r}{ r+  4\delta } \frac{1}{r-1} - \frac{1}{r-1}}{1 +  2\frac{1}{r-1} +  ( \frac{1}{r-1})^2}
\end{equation*}
which simplifies to
\begin{equation} \label{eq:final_difference_expression}
   \rho_{\text{star}} - \rho_{\text{complete}} = -\frac{4\delta(r-1)}{r(r+4\delta)}
\end{equation}
which holds for $\delta  \gg  \frac{1}{\sqrt{N}}$. Since $\rho_{\text{star}} - \rho_{\text{complete}} < 0 $ for all $\delta  \gg  \frac{1}{\sqrt{N}}$, the star is a suppressor. Therefore $\delta_c = \frac{1}{\sqrt{N}}$. 

\section{Calculating fixation probability numerically}
Gillespie algorithm is used to run the simulations following the method described in \cite{EGTMarkovContinuous}. 
\newline \newline 
To calculate fixation probability numerically, we note the total number of simulations run (reruns) $M_{\text{total}}$ and the number of those simulations that the mutant succeeded in, $M_\text{{mutant}}$. The fixation probability is then 
\begin{equation}
    \rho_{\text{numerical}} = \frac{M_\text{{mutant}}}{M_{\text{total}}}
\end{equation}
The more simulations are run, the closer the results gets to the analytical value. 

\section{Uncertainty in the simulation results seen in the Figures} \label{appendix:uncertainty_graphs}

The scatter points seen in most figures, ($M_{\text{total}}=10^4$ (\figurename\text{ }\ref{fig:contiuumBddB}, \figurename\text{ }\ref{fig:contiuumBddB_zoomzerotoone}, \figurename\text{ }\ref{fig:fixation_prob_difference_theory_sim}, \figurename\text{ }\ref{fig:fixation_prob_difference_theory_sim_error}, \figurename\text{ }\ref{fig:fixation_prob_difference_theory_sim_error_mean},\figurename\textbf{ }\ref{fig:contiuumBddB_zoomzerotoon_mean_error}) are the mean of five points of fixation probability calculated from 2000 simulations ($M_{\text{total}}=2 \times 10^3$). Hence the total number of simulations is $M_{\text{total}}=10^4$.  For \figurename\text{ }\ref{fig:discrete_star_complete_BddBupdate} $M_{\text{total}}=10^5$. This is is due to computational constraints. 
\newline \newline 
The error bars seen in \figurename\text{ }\ref{fig:contiuumBddB}, \figurename\text{ }\ref{fig:contiuumBddB_zoomzerotoone}, \figurename\text{ }\ref{fig:fixation_prob_difference_theory_sim_error} is the standard deviation of the five points used to estimate the mean, and is calculated by \cite{cumming2007error}
\begin{equation} \label{eq:standard_equation}
    s = \sqrt{\frac{1}{n}\sum_{i=1}^{n}(x_i - \bar{x})^2}
\end{equation}
where $x_i$ is the fixation probability calculated from 200 simulations (the measurement), $\bar{x}$ is the mean found from analytical equations (for star graph equation \ref{eq:star_fp_maintext} and for complete graph \ref{eq:complete_fp_maintext} found in section \ref{sec:continuous_time_model}), $n$ is the number of points used to estimate the mean (in this case 5).  
\newline \newline 
Using the standard deviation can calculate the standard error of the mean by \cite{cumming2007error}
\begin{equation} \label{eq:mean_error}
    s_n = \frac{s}{\sqrt{n}}
\end{equation}

\section{Additional Figures} \label{appendix:additional_graphs}
\begin{figure}
    \centering
    \includegraphics[scale=0.4]{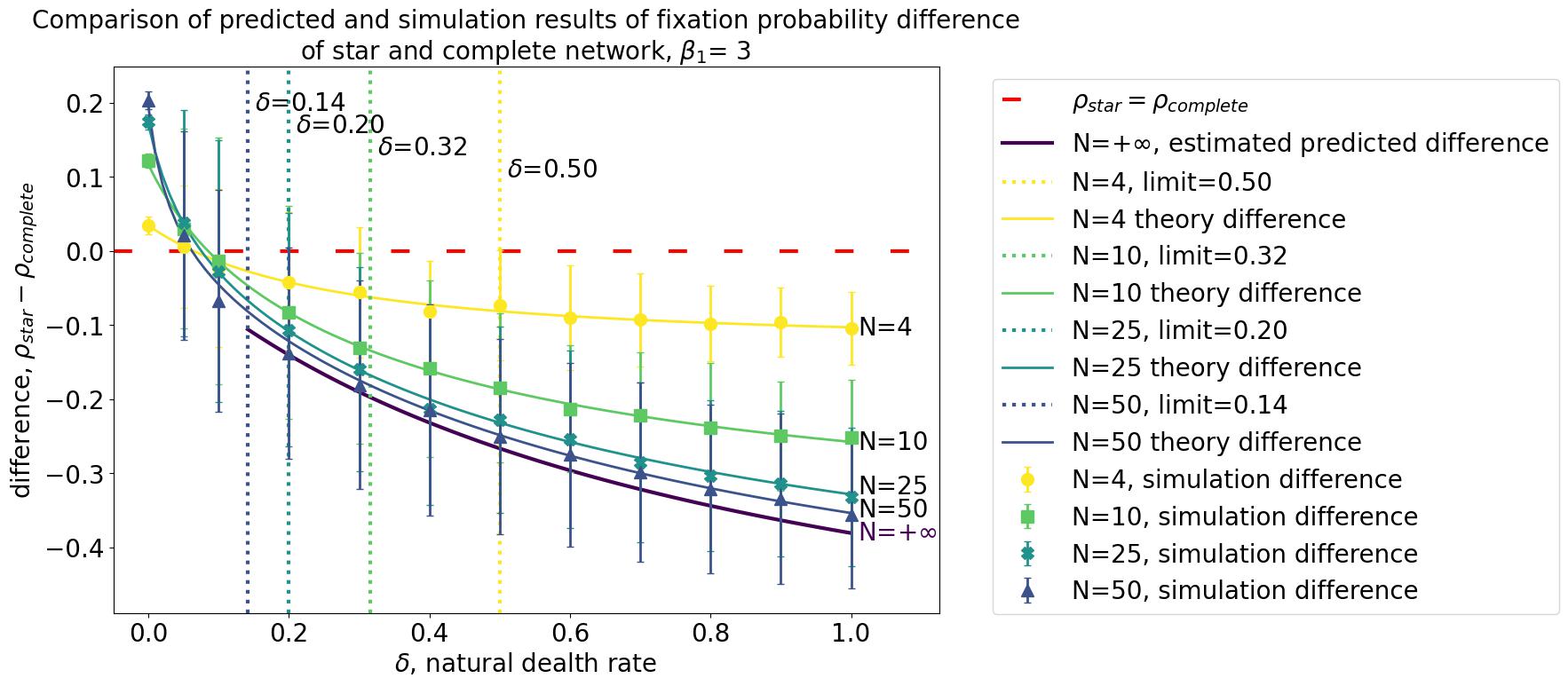}
    \caption{Showing how the difference of fixation probability of star and complete graph, $\rho_{\text{star}} - \rho_{\text{complete}}$, varies with the natural death rate $\delta$. This is the same figure as \figurename\textbf{ }\ref{fig:fixation_prob_difference_theory_sim} but with the standard deviation as the error bars, with method of calculating seen in section \ref{appendix:uncertainty_graphs}. The number of nodes $N=4,10,25,50$. The scatter points represent the numerical simulation results. 
    \newline 
    The solid lines represent the analytical results ($\rho_{\text{star}} - \rho_{\text{complete}}$, with $\rho_{\text{star}}$ defined by equation \ref{eq:star_fp_maintext} and $\rho_{\text{complete}}$ defined by equation \ref{eq:complete_fp_maintext}) for that size of graph (number of nodes). The horizontal dotted lines represent the upper bound limit of the change from amplifier to suppressor, as defined by equation \ref{eq:true_hold_maintext} and is dependent of $N$. The dashed green line represents the point of $\rho_{\text{star}} - \rho_{\text{complete}}=0$, which is important to note as this where the change from amplifier to suppressor occurs. 
    The other parameter values: the fitness of mutant $\beta_1=3$, offspring survival $s=1$, competition rate $\gamma=5$, amplifying rate $c=1000000$, the graph coefficients $w=w_c = \frac{1}{N}$ and $w_l=\frac{1}{2}$, reruns $10^4$.
     }
    \label{fig:fixation_prob_difference_theory_sim_error}
\end{figure}

\begin{figure}
    \centering
    \includegraphics[scale=0.4]{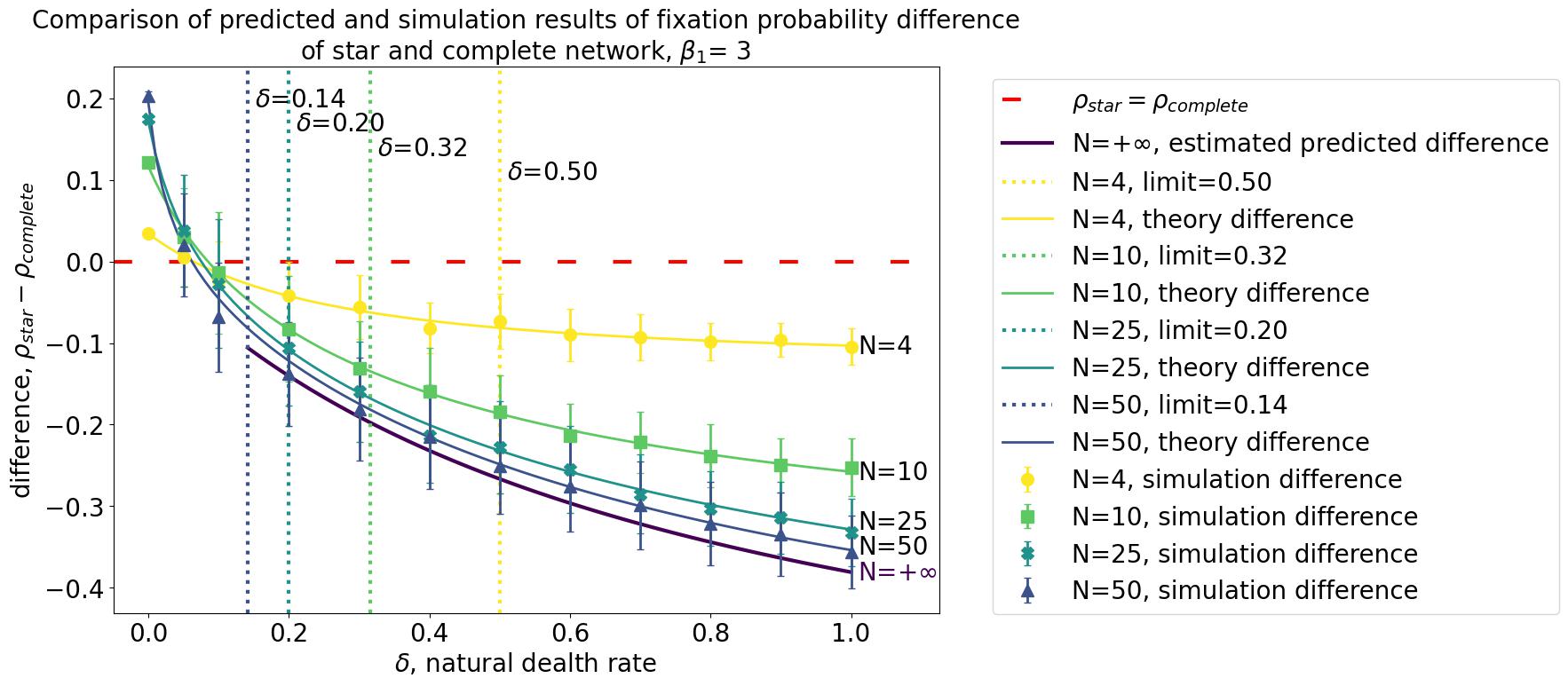}
    \caption{Showing how the difference of fixation probability of star and complete graph, $\rho_{\text{star}} - \rho_{\text{complete}}$, varies with the natural death rate $\delta$. This is the same figure as \figurename\textbf{ }\ref{fig:fixation_prob_difference_theory_sim} and \figurename\textbf{ }\ref{fig:fixation_prob_difference_theory_sim_error}, but with the mean error as the error bars, with method of calculating seen in section \ref{appendix:uncertainty_graphs}.
    \newline 
    The number of nodes $N=4,10,25,50$. The scatter points represent the numerical simulation results. 
    The solid lines of respective represent the analytical results ($\rho_{\text{star}} - \rho_{\text{complete}}$, with $\rho_{\text{star}}$ defined by equation \ref{eq:star_fp_maintext} and $\rho_{\text{complete}}$ defined by equation \ref{eq:complete_fp_maintext}) for that size of graph (number of nodes). The horizontal dotted lines represent the upper bound limit of the change from amplifier to suppressor, as defined by equation \ref{eq:true_hold_maintext} and is dependent of $N$. The dashed green line represents the point of $\rho_{\text{star}} - \rho_{\text{complete}}=0$, which is important to note as this where the change from amplifier to suppressor occurs. 
    The other parameter values: the fitness of mutant $\beta_1=3$, offspring survival $s=1$, competition rate $\gamma=5$, amplifying rate $c=1000000$, the graph coefficients $w=w_c = \frac{1}{N}$ and $w_l=\frac{1}{2}$, reruns $10^4$.
     }
    \label{fig:fixation_prob_difference_theory_sim_error_mean}
\end{figure}

\begin{figure}
    \centering
    \includegraphics[scale=0.4]{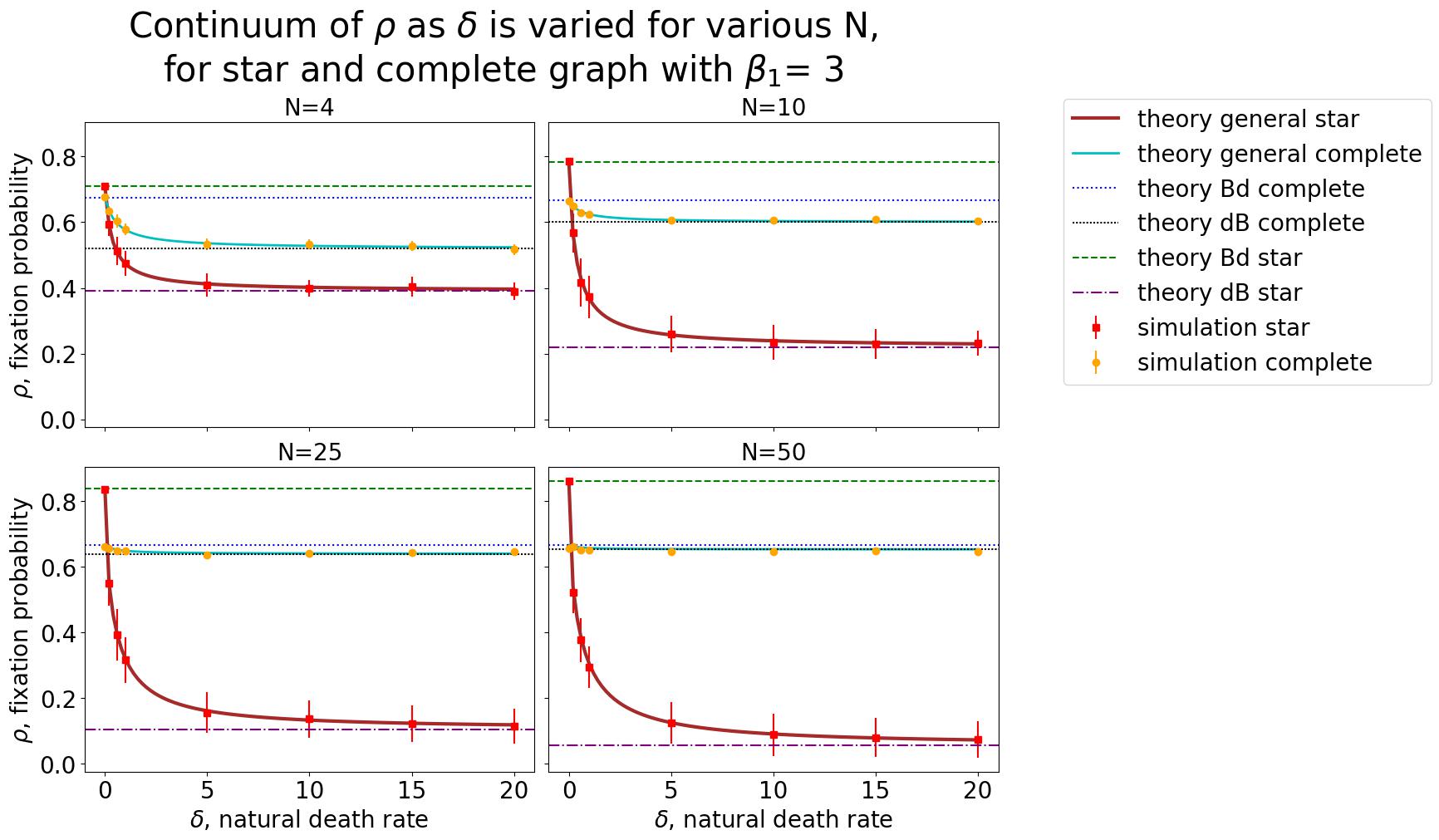}
    \caption{Showing how fixation probability $\rho$ varies with the natural death rate $\delta$ for complete and star graph. The fitness of the mutant is $\beta_1=3$, offspring survival rate is  $s=1$, number of nodes $N=4,10,25,40$, competition rates $\gamma=5$, amplifying rate $c=1000000$. The scatter points represents the mean of simulated results of $10^4$ simulations with the error bars being the standard error of the mean (method of calculating seen in section \ref{appendix:uncertainty_graphs}). The solid lines representing the general star and complete graph equations that were defined by equation \ref{eq:star_fp_maintext} and  \ref{eq:complete_fp_maintext} in section \ref{sec:continuous_time_model}. The horizontal lines are the analytical solutions of Bd and dB dynamics of the discrete-time model.
     }
    \label{fig:contiuumBddB_mean_error}
\end{figure}

\begin{figure}
    \centering
    \includegraphics[scale=0.4]{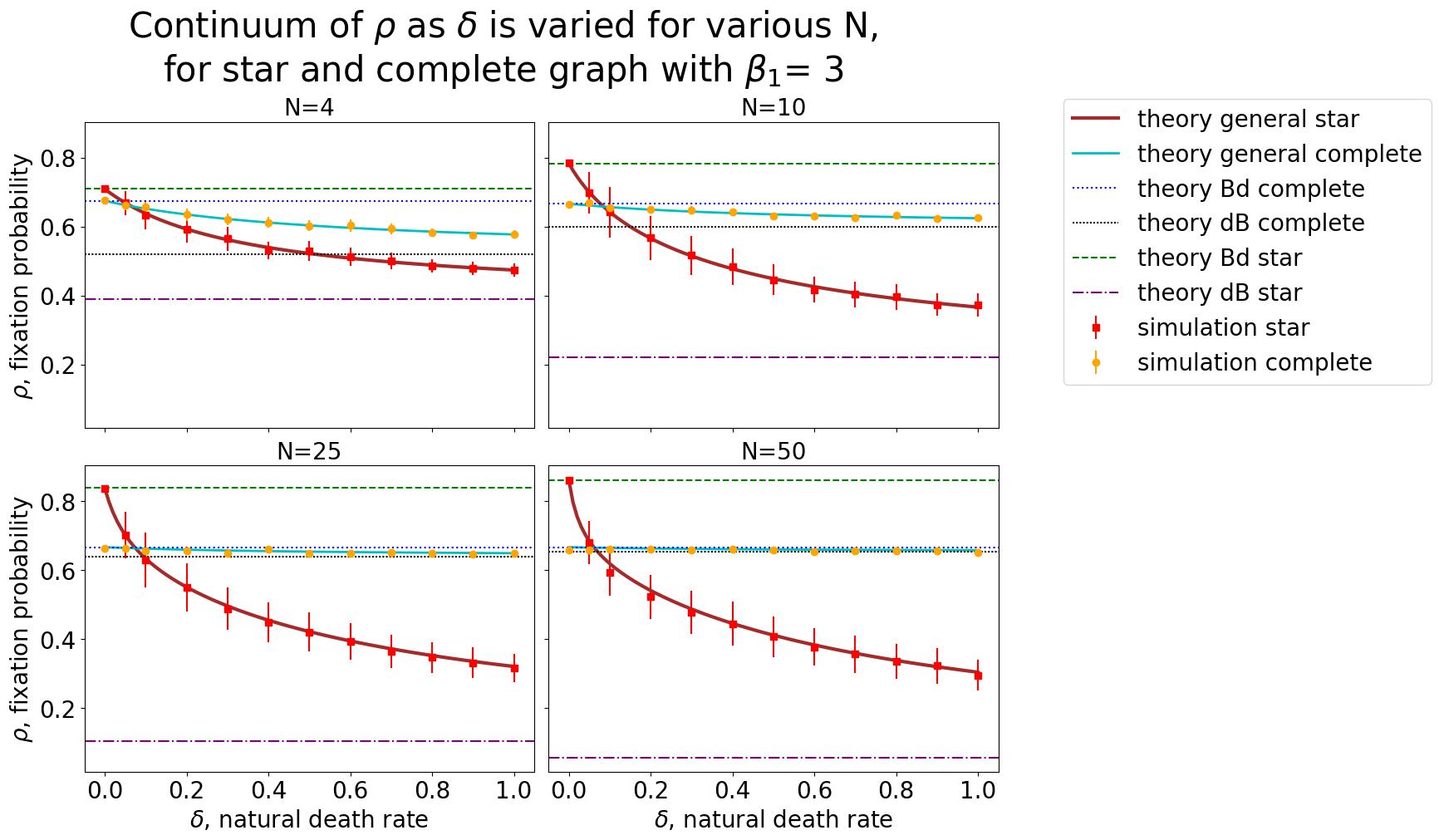}
    \caption{This shows the same results as in \figurename\textbf{ }\ref{fig:contiuumBddB} but focusing on range between of $\delta$ from 0 to 1. 
    As in \figurename\textbf{ }\ref{fig:contiuumBddB}, the fitness of the mutant is $\beta_1=3$, offspring survival rate is  $s=1$, number of nodes $N=4$, competition rates $\gamma=5$, amplifying rate $c=1000000$. The scatter points represents the mean of simulated results of $10^4$ simulations with the error bars being the standard error of the mean (method of calculating seen in section \ref{appendix:uncertainty_graphs}). The solid lines representing the general star and complete graph equations that were defined by equation \ref{eq:star_fp_maintext} and  \ref{eq:complete_fp_maintext} in section \ref{sec:continuous_time_model}. The horizontal lines are the analytical solutions of Bd and dB dynamics of the discrete-time model. \\ 
     }
    \label{fig:contiuumBddB_zoomzerotoon_mean_error}
\end{figure}

For reference, here is the \figurename\textbf{ }\ref{fig:fixation_prob_difference_theory_sim_error} with the standard deviation as the error bars and \figurename\textbf{ }\ref{fig:fixation_prob_difference_theory_sim_error_mean}, \figurename\textbf{ }\ref{fig:contiuumBddB_mean_error}, \figurename\textbf{ }\ref{fig:contiuumBddB_zoomzerotoon_mean_error} with the standard error of the mean.

\newpage 

\clearpage \newpage  


\end{appendices}
\end{document}